\documentclass[preprint2]{aastex}

\shorttitle{HD 46487 is now a Be Star}
\shortauthors{Whelan \& Baker}

\begin{document}

\title{HD~46487 is Now a Classical B$e$\ Star}

\author{David G. Whelan}
\affil{Department of Physics, Austin College}
\email{dwhelan@austincollege.edu}

\and

\author{R. David Baker}
\affil{Department of Physics, Austin College}

% AAVSO keywords = Spectra, Spectroscopy; Variable Star Observing; Variable Stars (Individual); HD 46487; Spectroscopic Analysis; Be Stars

\begin{abstract}

We present the first observations of hydrogen line emission detected
around the B-type star HD~46487, a well-studied star in the CoRoT
field of view. The emission is only evident in the H$\alpha$\ line,
for which the observed violet-red peak separation ($\Delta v_{p}$) is
typical of a B$e$\ star with a circumstellar disk. The absence of dust
emission from the infrared spectral energy distribution excludes the
possibility of a very young star. The star's magnitude (V$=5.079$) and
regular use in the literature for a variety of studies suggests that
the line emission had a high probability of being found previously,
had it been evident; since such was not the case, we believe that the
B$e$\ phenomenon for HD~46487 has only very recently ``turned on.'' We
therefore recommend that this star be spectroscopically and
photometrically monitored to track continued changes to its
circumstellar morphology.

\end{abstract}


\section{Introduction}\label{intro}

Main sequence and giant B-type stars are often fast
rotators. Classical B$e$\ stars are universally so, and most are
believed to rotate at speeds near critical \citep{Townsend04}. Unlike
``normal'' B-type stars, however, B$e$\ stars experience nonradial
pulsational modes that eject matter from the surface, allowing them to
form a circumstellar decretion disk at the star's equator
\citep{Rivinius13}. It is therefore interesting that stars have become
B$e$\ stars after decades of appearing normal, or else have ceased to
exhibit the B$e$\ phenomenon \citep{Chojnowski15,Chojnowski17}. In
fact, observations of the transition between normal B-type and
B$e$\ star are becoming quite common; many examples exist in the more
than decade-long archive of the B$e$\ Star Spectra (BeSS)
Database\footnote[1]{BeSS may be accessed online at
  http://basebe.obspm.fr/basebe/Accueil.php?flag\_lang=en}
\citep{Neiner11}. With the help of BeSS, data are being collected that
illustrate the timescales on which disks build up and decrete, track
violet-to-red emission peak separation (VR) variability due to
one-armed global oscillations \citep{Okazaki91}, and variability due
to outbursts.

But in spite of this growing collection of data, little is known about
why they begin to exhibit emission when they do
\citep{Porter03}. \citet{McSwain09} showed that B$e$\ stars rotate
faster on average than normal B-type stars, and their near-critical
speeds make it easier for B$e$\ stars to lose mass to their
surroundings. The difference in rotation rate for roughly 75\%\ of
B$e$\ stars is likely due to the transfer of angular momentum during
binary interactions \citep{McSwain05, deMink13}, and evidence suggests
that such interactions are most common within the first 100~Myr of the
star's life.

HD~46487, unlike many recently discovered B$e$\ stars, has been
well-studied of late; partly this is because it is a bright source
(V$=5.079$) but also because it resides in the CoRoT field of view,
which was targeted for exoplanet discoveries and asterseismology
studies \citet{Auvergne09}. At the time that the CoRoT field of view
was being searched for B$e$\ stars, HD~46487 was not one, and so was
not included in \citet{Neiner05} or \citet{Fremat06}. However, other
publications from this time allow us to search for any evidence of
extant or forming circumstellar matter in the past couple of
decades. Its far-ultraviolet (FUV) spectrum exhibited no resonance
lines in 2003, as we might have expected for a B$e$\ star \citep{Jo16,
  Rountree91}. It also showed no evidence of any photospheric
pulsation \citep{Lefever10}; photometric variability, due to either
periodic variability or outbursts, is a common feature of B$e$\ stars
\citep{Labadie16,Rivinius13b}. Other more recent uses for HD~46487 in
the literature, such as being used as a calibrator in interferometric
observations \citep[{\it e.g.},][]{Ellerbroek15}, further suggest that
the star was still normal until as late as 2013.

Several studies have provided us with fundamental physical
parameters. HD~46487 was classified as a B5~Vn star in \citet{Abt90},
where the ``n'' designation suggests broad absorption lines. Its
projected rotational speed, $v\sin i$\ where $i$\ is the inclination
angle, is likely between 285-300~km/s \citep{Abt02,Huang10}. This high
speed has lead to a bulging out of the equator, such that the equator
has a substantially lower surface gravity than its poles
\citep[log($g$) of 3.63 at the equator, compared to 3.95 at the
  poles;][]{Huang10}. There is no observed binary companion
\citep{Abt90, Eggleton08,Gullikson16a}. Recent spectra of HD~46487 are
reported in \citet{Gullikson16a,Gullikson16b}. Presented in these
works there is a high-resolution (R$\sim$60,000) spectrum in the
wavelength range 3400\AA -10000\AA\ taken with the cross-dispersed
echelle spectrograph (TS23) at the Harlan J. Smith 2.7-meter
telescope\footnote[2]{http://www.as.utexas.edu/mcdonald/facilities/2.7m/cs2.html}
on 2014-01-11, as well as two near-infrared
(1.45$\micron$-2.5$\micron$) spectra taken with IGRINS \citep{Park14}
on the same telescope taken on 2014-10-16 and 2015-03-03. These three
spectra together cover wavelengths for Balmer, Paschen, and Brackett
series hydrogen lines, and would all have clearly exhibited that
HD~46487 was an emission-line source, had it been evident.

We present the first known observations, taken in March 2017, of the
onset of the B$e$\ phenomenon in the well-studied star HD~46487. We
will describe our observations and data reduction in
Section~\ref{obs}, present and analyze our spectra in
Section~\ref{analysis}, briefly discuss the results in
Section~\ref{discussion}, and then conclude in
Section~\ref{conclusion}.

%\begin{center}
%  \begin{deluxetable}{cccccc}
%    \tabletypesize{\footnotesize}
%    \tablecaption{HD 46487 Observations Data\label{obslog}}
%    \tablehead{MJD & Dispersion & Wavelength Range & Exposure Time & Airmass & Average SNR \\  & (\AA/pixel) & (\AA) & (seconds) & & }
%    \startdata
%    57826.098 & 0.168 & 6380-6722 & 1800 & 1.3 & 196 \\
%    57827.054 & 0.168 & 6380-6722 & 1800 & 1.2 & 151 \\
%    57827.103 & 0.235 & 4182-4673 & 1800 & 1.3 & 339 \\
%    57827.141 & 0.543 & 3820-4942 &  900 & 1.4 & 304 \\
%    \enddata
%  \end{deluxetable}
%\end{center}

%\clearpage

\section{Observations and data reduction}\label{obs}

The Adams Observatory sits atop the IDEA Center science building at
Austin College in Sherman, TX. This facility provides opportunities
for research, introductory and advanced astronomy classes, and public
star-gazing events. Built by DFM Engineering in 2013, the
0.61-m $f/8$\ Ritchey-Chr\'{e}tien telescope is used primarily for
spectroscopy, photometry, and imaging. Instruments are located at
Cassegrain focus.

The spectrograph used for these observations is a long-slit LhiresIII
spectrograph designed for commercial sale by Shelyak Instruments. It
is a modular spectrograph for which the dispersion grating can easily
be switched. Collimation and focusing are performed by the same optic,
a simple doublet with $f/6.67$\ and a diameter of 30~mm.

The CCD camera being used is a Finger Lakes Instrumentation (FLI)
Microline with a thermoelectric cooler that can reach $60\degr$~C
below ambient. It contains a back-illuminated e2V 42-10 CCD that is
coated for enhanced broadband transmittance ($\sim$75-95\% quantum
efficiency from $\lambda = 3800$\AA - 7000\AA) and the array of
512$\times$2048 pixels are each 13.5$\micron$\ square. Considering the
demagnification of the spectrograph camera, the effective pixel size
at the slit mask is 16$\micron$. We therefore choose a
35$\micron$\ slit as the best resolution match (a 32$\micron$\ slit is
not available for sale through Shelyak). The camera and spectrograph
can be used to observe stars as faint as $V \sim 10$.

Dispersion with the 2400 gr/mm dispersion grating is 0.168\AA~per
pixel around 6500\AA, and 0.235\AA~per pixel around 4300\AA. The
resultant resolutions vary with wavelength. Typically we see
resolutions in the range $7,500 \lesssim R \lesssim 9,500$ around
H$\gamma$ (4341\AA) and $14,000 \lesssim R \lesssim 20,000$ around
H$\alpha$ (6563\AA). With the 1200 gr/mm grating, dispersion is
0.54\AA~per pixel in between $\sim 3800$\AA\ and $5000$\AA, with the
resolution varying between $3,000 \lesssim R \lesssim 4,500$.

A log of observational data, including exposure times and
signal-to-noise ratios (SNR) is provided in Table~1. HD~46487 was
first observed as a telluric standard for another project, so only a
single spectrum was acquired on the first night. Flatfield, dark
current, bias, and neon-argon lamp images are observed every
night. The gain and read noise are computed using the bias and
flatfield images; gain is 1.5~$e^{-}/$ADU and read noise is
13.1~$e^{-}/$pixel. Dark current and bias are removed from the science
images, and they are then divided by the normalized flatfield. All
data reduction and spectral extraction is performed in {\sc PYTHON}
using the author's own routines.

%\begin{center}
%  \begin{deluxetable*}{cccccc}
%    \tabletypesize{\footnotesize}
%    \tablecaption{HD 46487 Observations Data\label{obslog}}
%    \tablehead{MJD & Dispersion & Wavelength Range & Exposure Time & Airmass & Average SNR \\  & (\AA/pixel) & (\AA) & (seconds) & & }
%    \startdata
%    57826.098 & 0.168 & 6380-6722 & 1800 & 1.3 & 196 \\
%    57827.054 & 0.168 & 6380-6722 & 1800 & 1.2 & 151 \\
%    57827.103 & 0.235 & 4182-4673 & 1800 & 1.3 & 339 \\
%    57827.141 & 0.543 & 3820-4942 &  900 & 1.4 & 304 \\
%    \enddata
%  \end{deluxetable*}
%\end{center}

\begin{center}
  \begin{tabular}{c c c c c c}
    \multicolumn{6}{c}{Table 1. HD 46487 Observations Data\label{obslog}} \\
    MJD & Disp.  & $\lambda$~Range & Exp $t$ & sec Z & SNR \\
    & (\AA/pix) & (\AA)            & (sec)     &         &             \\
    \hline
    57826.10 & 0.168 & 6380-6722 & 1800 & 1.3 & 196 \\
    57827.05 & 0.168 & 6380-6722 & 1800 & 1.2 & 151 \\
    57827.10 & 0.235 & 4182-4673 & 1800 & 1.3 & 339 \\
    57827.14 & 0.543 & 3820-4942 &  900 & 1.4 & 304 \\
    \hline
  \end{tabular}
\end{center}

\vspace{1mm}

The reduced science images are collapsed in the wavelength direction,
and the star's dispersion is fit with a Gaussian to determine an
extraction center and width. The spectrum is then extracted between
2$\sigma$\ of the Gaussian's center, rounded out to the nearest
pixel. The pixels on either side of the spectrum's extraction window
are used to compute a local sky value that is then subtracted away
from the star's spectrum, removing any background. Wavelength values
are determined using known lines from the neon-argon lamp
spectrum. The SNR is computed at every pixel using the CCD equation
from \citet{MerlineHowell95}, and are then propagated to compute
errors at every wavelength value in the spectrum.

\section{Analysis}\label{analysis}

\subsection{Disk emission}\label{rd}

\begin{figure*}
  \includegraphics[scale=0.70]{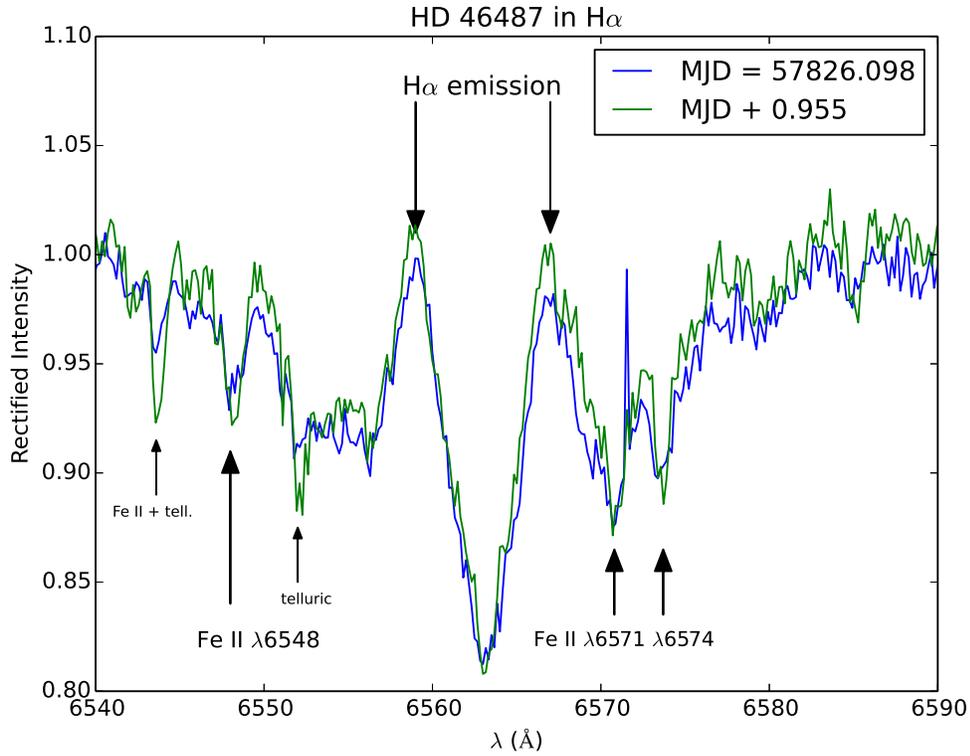}
  \caption{Two spectra observed on consecutive nights show
    H$\alpha$\ emission inside the photospheric absorption line. Various
    Fe~{\sc ii} lines are evident, as well as some telluric absorption
    lines. Modified Julian dates are shown.\label{Ha}}
\end{figure*}

The H$\alpha$\ spectrum of HD~46487 is shown in Figure~\ref{Ha}, as it
was observed on the nights of 13 and 14 March, 2017. There are
noticeable differences between the two spectra. The telluric
absorption features are substantially stronger in the second
observation. The hydrogen emission is markedly more pronounced as
well. Variations in emission strength on the order of days is not
uncommon for B$e$\ stars, particularly during a phase of disk build-up
\citep[{\it e.g.},][and references therein]{Rivinius13}.

The peak separation in the H$\alpha$\ emission is $\sim$7.9\AA, which
translates to an orbital speed of $\sim$180~km/s. If we take $v\sin i
= $ 290~km/s as a rough average of the available literature values
(see Section~\ref{intro} above), then according to Huang's Law
\citep{Huang72}:

\begin{equation}
r_d = \left(\frac{2\times v \sin i}{\Delta v_p}\right)^2
\end{equation}

\noindent where $\Delta v_p$ is the emission peak separation, then the
H$\alpha$\ line-emitting radius is 19 $R_{*}$.

\subsection{Spectral typing}\label{spt}

\begin{figure*}
\includegraphics[scale=0.32]{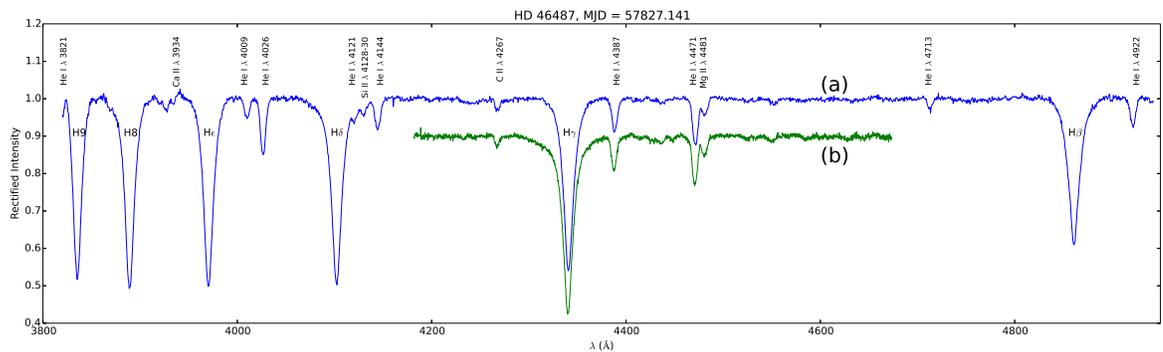}
\caption{The violet-blue-green spectrum of HD~46487 taken on the night
  of 14 March 2017, with hydrogen, He~{\sc i} and various metal
  absorption lines labeled. Spectrum (a) was obtained using the
  1200~gr/mm dispersion grating, (R~$\sim 4,000$) while spectrum (b)
  was obtained using the 2400~gr/mm grating (R~$\sim
  9,000$).\label{uvvis}}
\end{figure*}

The optical spectrum of HD~46487 in the wavelength range
3820\AA~-~4950\AA\ is shown in Figure~\ref{uvvis} at two
resolutions. The strengths of the He~{\sc i} $\lambda 4009/4026$
lines, the presence of Si~{\sc ii} $\lambda 4128-4130$, its strength
relative to He~{\sc i} $\lambda 4121, 4144$, and the He~{\sc i}
$\lambda 4471$/Mg~{\sc ii} $\lambda 4481$ ratio all confirm a spectral
type of B5. The lines are very broad, suggesting that this is a main
sequence star, and that the ``nebular'' designation is justified. We
conclude that the spectral type based on visual inspection is B5~Vn,
and that emission is not evident in the optical spectrum.

We measured the equivalent widths for the helium and metal absorption
lines observed in the optical spectra. Results are given in Table~2,
along with plane-parallel model predictions from \citet{Fremat06} for
comparison. Most notably, all of the line equivalent widths would
suggest an earlier spectral type than what is visually observed.

\begin{center}
  \begin{tabular}{l c c c c}
    \multicolumn{5}{c}{Table 2. Equivalent Width Measurements\label{ews}} \\
    Line ID & \multicolumn{4}{c}{Equivalent Widths (m\AA)} \\ 
    & Low-res\tablenotemark{a} & Med-res\tablenotemark{b} & B2V\tablenotemark{c} & B5V\tablenotemark{c} \\
    \hline
    He~{\sc i}$\lambda 4009$ &  $324 \pm 19$ &  --           &  613 & 217 \\
    He~{\sc i}$\lambda 4026$ &  $977 \pm 20$ &  --           & 1541 & 878 \\
    He~{\sc i}$\lambda 4144$ &  $524 \pm 20$ &  --           &  765 & 322 \\
    C~{\sc ii}$\lambda 4267$ &  $144 \pm 13$ &  $145 \pm 17$ &  270 &  97 \\
    He~{\sc i}$\lambda 4387$ &  $612 \pm 16$ &  $563 \pm 23$ &  950 & 378 \\
    He~{\sc i}$\lambda 4471$ &  $683 \pm 33$ &  $742 \pm 34$ & 1442 & 667 \\
    Mg~{\sc ii}$\lambda 4481$&  $193 \pm 33$ &  $225 \pm 30$ &  198 & 272 \\
    \hline
    \multicolumn{5}{l}{$^a$\ Data taken with the 1200 gr/mm dispersion grating.}\\
    \multicolumn{5}{l}{$^b$\ Data taken with the 2400 gr/mm dispersion grating.}\\
    \multicolumn{5}{l}{$^c$\ Model values from Fr\'{e}mat et al.(2006).}\\
  \end{tabular}
\end{center}

%\begin{center}
%  \begin{deluxetable}{ccccc}
%    \tabletypesize{\footnotesize}
%    \tablecaption{Equivalent Width Measurements\label{ews}}
%    \tablehead{Line ID & \multicolumn{4}{c}{Equivalent Widths (m\AA)} \\ & Low-res\tablenotemark{a} & Med-res\tablenotemark{b} & B2V\tablenotemark{c} & B5V\tablenotemark{c} }
%    \startdata
%    \hline
%    He~{\sc i}$\lambda 4009$ &  $324 \pm 19$ &  --           &  613 & 217 \\
%    He~{\sc i}$\lambda 4026$ &  $977 \pm 20$ &  --           & 1541 & 878 \\
%    He~{\sc i}$\lambda 4144$ &  $524 \pm 20$ &  --           &  765 & 322 \\
%    C~{\sc ii}$\lambda 4267$ &  $144 \pm 13$ &  $145 \pm 17$ &  270 &  97 \\
%    He~{\sc i}$\lambda 4387$ &  $612 \pm 16$ &  $563 \pm 23$ &  950 & 378 \\
%    He~{\sc i}$\lambda 4471$ &  $683 \pm 33$ &  $742 \pm 34$ & 1442 & 667 \\
%    Mg~{\sc ii}$\lambda 4481$&  $193 \pm 33$ &  $225 \pm 30$ &  198 & 272 \\
%    \hline
%    \enddata
%    \tablenotetext{a}{Data taken with the 1200 gr/mm dispersion grating}
%    \tablenotetext{b}{Data taken with the 2400 gr/mm dispersion grating}
%    \tablenotetext{c}{Model values from Fr\'{e}mat et al.(2006)}
%  \end{deluxetable}
%\end{center}

\subsection{The spectral energy distribution}\label{ir}

\begin{figure*}
\includegraphics[scale=0.45]{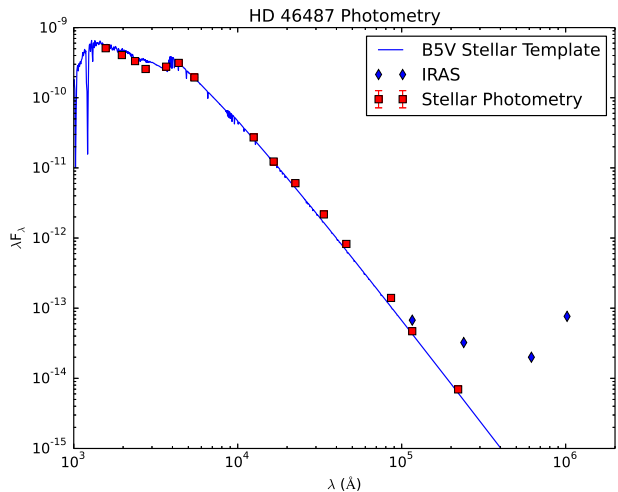}
\caption{The infrared SED for HD~46487 is plotted against a B5~V
  stellar template for comparison. IRAS data points (diamonds) have a
  point spread function that includes nearby diffuse dust
  emission. Error bars are smaller than the symbol
  size.\label{irphot}}
\end{figure*}

The spectral energy distribution (SED) from ultraviolet to infrared is
plotted against a B5~V stellar template in Figure~\ref{irphot}. The
stellar template is taken from \citet{CastelliKurucz}. Data points
were collected using VizieR and sources include \citet{Thompson78} for
the ultraviolet, \citet{Crawford71} and \citet{Hog00} for the optical,
and the 2MASS survey \citep{Skrutskie06}, the WISE Survey
\citep{Wright10}, the AKARI All-Sky Survey \citep{Ishihara10}, and the
IRAS survey \citep{Neugebauer84} for the infrared. The IRAS data
points suffer from a very large point spread function ($\sim
5\arcmin$) and it is no surprise that they contain emission from
nearby interstellar gas; this is the source of the discrepancy between
the IRAS and WISE data points. There is no evidence of dust emission
around HD~46487, and out to 22~\micron, there is no substantial
deviation from the stellar Rayleigh-Jeans tail.

\section{Discussion}\label{discussion}

The H$\alpha$\ line-emitting radius is calculated to be 19~R$_{*}$, as
discussed in Section~\ref{rd}. This is consistent with the average
H$\alpha$\ line-emitting radii for B$e$\ stars in \citet{Hanuschik88}
and in \citet{Slettebak92}, which quote $\sim~20$~R$_{*}$\ and
$\sim~19$~R$_{*}$ respectively.

The spectral type determined upon visual inspection should be treated
with skepticism. Fast-rotating stars bulge at the equator, which both
increases their surface area and creates a gradient in surface
temperature from equator to pole. As a result, fast-rotating stars
have higher luminosities and lower average surface temperatures ({\it
  i.e.}, later spectral types) than their slow-rotating counterparts
\citep{Gray09}. There is also the effect of rotation on the perceived
depths of the absorption lines themselves. Fast rotation will broaden
the helium and metal absorption lines, so that they appear shallower
than they would for another star of the same spectral type. These
broadening effects are, in some cases, asymmetric, as is the case with
the Mg~{\sc ii}~$\lambda 4481$\ line, which is intrinsically narrower
than the He~{\sc i}~$\lambda 4471$\ line. The Mg~{\sc ii}~$\lambda
4481$/He~{\sc i}~$\lambda 4471$\ line ratio is one of several used in
spectral typing for which intrinsic line width differences can be an
issue. Since issues related to rapid rotation affect our perception of
a star's intrinsic spectral type, some work has been done to spectral
type fast-rotating stars \citep[{\it e.g.},][]{Garrison94}. At this
time, however, there exist no fast-rotating spectroscopic standards
earlier than B7 in the literature that can be used for visual
comparisons.

The spectral type inferred from equivalent width measurements can also
be problematic for B$e$\ stars. Continuum emission originating within
the disk can partially fill in the absorption lines in a process known
as line damping. This means that a B$e$\ star's actual spectral type
will be earlier than that measured. This is in addition to the effects
of scattered light from within the spectrograph itself, which will
also fill in absorption signatures. Since the equivalent width
measurements in Table~2 already suggest that the spectral type
is earlier than what is inferred from visual inspection, we may
consider line damping and scattered light as exaggerating effects, and
can confidently conclude that HD~46487 has an earlier spectral type
(B4 or B3) than is determined from visual inspection.

Pre-main sequence stars such as Herbig B$e$\ stars exhibit hydrogen
line emission due to a circumstellar disk just like classical
B$e$\ stars \citep{Herbig60}. They should additionally possess a broad
infrared excess due to circumstellar free-free emission \citep[which
  is expected for classical B$e$\ stars as well;][]{Gehrz74} and
thermally-radiating dust \citep{Malfait98}. Indeed, \citet{McDonald12}
found 1~magnitude of infrared excess for HD~46487 out to
22~$\micron$. As illustrated in Figure~\ref{irphot}, we cannot confirm
such a large infrared excess, and what little infrared excess is seen
is certainly not due to dust emission. HD~46487 is therefore not a
pre-main sequence source, and what infrared excess exists is most
likely due to circumstellar gaseous material and/or winds that would
be responsible for free-free emission.

Due to the recent IGRINS and TS23 spectra of HD~46487 published in
\citet{Gullikson16a}, we can state that HD~46487 was observed to be a
normal main sequence B-type star as late as March 2015. All of the
data used to compile the SED in Figure~\ref{irphot} was published
earlier than 2013. It was additionally used in 2013 as a calibrator
for interferometric observations in
Br$\gamma$\ \citep[][]{Ellerbroek15}, an unsatisfactory choice had it
exhibited a substantial gaseous circumstellar disk. It therefore seems
likely that the B$e$\ phenomenon became evident no earlier than March
2015.

\section{Conclusion}\label{conclusion}

HD~46487 (V$=5.079$) makes an excellent target for small aperture
telescopes. Now that it exhibits the B$e$\ phenomenon, we should
expect it to vary like other B$e$\ stars: with periodic photometric
variability, the occasional small outburst, and variations in its line
emission on timescales anywhere from hours to years. There are three
specific areas that will be most useful for continued studies of this
source.

\begin{itemize}
\item {\bf Spectroscopic observations of H$\alpha$.} Such observations
  would be useful to track the changing line emission, whether it be
  due to small outbursts and resultant pockets of gas rotating within
  the disk, or else long-scale global one-armed oscillations. Spectra
  submitted to BeSS (as we plan to do) would then be available to the
  entire community of B$e$\ star observers.
\item {\bf Photometric observations at optical wavelengths.} Such
  observations may be useful for tracking periodicity, which is common
  in B$e$\ star atmospheres, and may also be useful for catching
  outbursts.
\item {\bf Near-infrared photometric observations.} Most B$e$\ stars
  show an infrared excess due to free-free emission. Now that HD~46487
  exhibits the B$e$\ phenomenon, we expect to see this change mostly
  to its near-infrared SED.
\end{itemize}

We recognize the need for collaboration on these
observations. Interested parties are encouraged to contact the authors.

\acknowledgements 

This work would not have been possible without the Adams Observatory
at Austin College, made possible by the generosity of the John and
Patricia Adams Foundation of Bedford, Texas. DGW would like to thank
Michael Joner for his critical read of the manuscript before
submission, and for the helpful feedback of the anonymous referee.

\end{document}